# Physical Consequences of a Special Conformal Invariance of Maxwell's Equations

Carl E. Wulfman[+]
Department of Physics, University of the Pacific
Stockton, CA

The velocity of light is invariant under transformations that alter spacetime metrics, while leaving Maxwell's equations invariant. A one-parameter special conformal invariance group of the equations alters the standard Minkowski metric and exposes an ambiguity in current interpretations of the Doppler effect. Comparisons between Doppler measurements and direct ranging measurements of velocities and positions of distant spacecraft could determine the value of the group parameter. The metric is a conformal metric and not a Minkowski metric, if the group parameter is found to be non-zero. In this case, current understandings of the physics of EM wave transmission, the Doppler effect, and Hubble's relations, must be substantially revised.

[+] Professor Emeritus. Email: wulfmanc@wavecable.com

**Introduction**

In 1908 Minkowski introduced a spacetime metric that is invariant under Lorentz and Poincare transformations.[1] A year later Bateman and Cunningham proved that Maxwell's equations, and the velocity of light in vacuo, are invariant under the transformations of a much larger, conformal group.[2] The group contains a discrete inversion transformation, and a fifteen-parameter Lie group containing the ten-parameter Poincare group. The Minkowski metric is altered by the remaining transformations of this Lie group, and by the inversion transformation. Bateman and Cunningham used Minkowski's transformation, t -> it, to obtain their group from the inversion transformation of the conformal group $C^{(4)}$ investigated by Lie.[3] The resulting inversion through the origin of a light cone transforms the generators of translations with respect to spacetime coordinates, to four generators, $C_a$, of special conformal transformations. Let c represent the speed of light, t represent time, and let $x^1, x^2, x^3, x^4 = ct$, be cartesian coordinates with origin at the center of the light cone. Also let $r = |\mathbf{r}|$, $\mathbf{r} = (x^1, x^2, x^3)$. Then the generator of time translations is converted into

$$C_4 = (r^2 + (x^4)^2)/x^4 + 2 x^4 \, r/r. \tag{1}$$

$C_4$ generates a one-parameter Lie group of transformations of $|\mathbf{r}|$, and $x^4$ in the four-vector, $\mathbf{x} = (\mathbf{r}, x^4)$. $C_4$ is invariant under space reflections, and like $\partial/\partial t$, changes sign under time reflection. The origin is an invariant point of the transformations generated by $C_4$. The four operators $C_a$, together with the spacetime dilatation operator and the generators of the Poincare group, generate the fifteen-parameter Bateman-Cunningham conformal Lie group which acts on these coordinates. The group is locally isomorphic to SO(4,2).[4]

Just as World War II ended, E. L. Hill[5] pointed out that the conformal invariance transformation of Maxwell's equations generated by $C_4$ establishes a relation isomorphic to Hubble's laws, which seemed to imply that Hubble's laws had a purely kinematic foundation. Our analysis is prompted by Hill's startling discovery, but we do not assume *a priori* that it has this implication. In the discussion below we develop the consequences of finite transformations generated by $C_4$. It is shown that if the group parameter is nonzero, the resulting spacetime metric alters the wavelength of EM waves presupposed measured in Minkowski space. This changes the interpretation of observed Doppler shifts. It is then shown that the value of the group parameter can be experimentally established. Its value provides a measure of the extent to which Hubble's law is a consequence of dynamics, and the extent to which it is simply a consequence of kinematics describing motions in the conformal spacetime.

Hill's much overlooked paper is mentioned in Kastrup's recent review of the history and applications of conformal groups,[6] but in so far as we are aware, there have until now been no further investigations of Hill's discovery. However a recent paper by Tomilchik[7] develops a more general analysis that utilizes transformations of the Bateman-Cunningham conformal group, and concludes, as Hill apparently did, that Hubble's law has a purely kinematic foundation. A number of authors have used conformal groups in general-relativistic treatments of gravitationally produced red shifts. In this connection, the reader is referred to the book of Hoyle, Burbidge, and Narlikar,[8] and to recent articles by Chodorowski[9], and by Bunn and Hogg[10].

## 2. The Conformal Transformation Arising From Time Translations

The operator $\exp(\beta_4 C_4)$ produces the finite transformations[4]

$$r \to r' = r, \quad x^4 \to x'^4 = \gamma\{x^4 - \beta_4 s^2\}$$

in which                                                                                                              (2a-e)

$$s^2 = (x^{4\,2} - r^2), \quad \gamma = \gamma(\beta_4, x^4, r) = (1 - 2\beta_4 x^4 + (\beta_4)^2 s^2)^{-1}.$$

It leaves invariant any function of $(x^{4\,2} - r^2)/r = s^2/r$, so equations (2) imply that

$$\exp(\beta_4 C_4) s^2 = \gamma s^2. \tag{3a}$$

Using them, one finds

$$dr \to dr' = \gamma^2 (A\, dr + B\, dx^4),$$

and (3b,c)

$$dx^4 \rightarrow dx'^4 = \gamma^2 (B\, dr + A\, dx^4),$$

with

$$A = 1 - 2\beta_4 x^4 + \beta_4^2(r^2 + x^{4\,2}), \quad B = 2\beta_4 r(1 - \beta_4 x^4). \tag{3d,e}$$

Thus

$$Dr'/dx'^4 = (A\, dr/dx^4 + B)/(B\, dr/dx^4 + A). \tag{3f}$$

We will consider these transformations to be mappings relating the two sets of coordinates, $(r,t,v = dr/dt)$ and $(r',t',v' = dr'/dt')$, of a point in a phase space which accommodates particles of zero rest-mass. Setting $\beta_4 = \alpha/2c$, and expanding (3b-e) to first order in $\alpha$, one obtains Hill's relations:

$$R' = (1 + \alpha t)r, \quad t' = t + \alpha(r^2/c^2 + t^2)/2,$$
$$dr' = dr(1 + \alpha t) + \alpha r\, dt, \quad dt' = dt(1 + \alpha t) + \alpha r\, dr/c^2, \tag{4a-e}$$
$$v' = v + \alpha r (1 - v^2/c^2).$$

Setting $\alpha = -\alpha$, and interchanging primed, unprimed, symbols gives their inverse.

The finite transformations set forth in (3) change the Minkowski metric

$$ds^2 = |(\mathbf{dr})^2\ (dx^4)^2|, \tag{5a}$$

to the conformal metric

$$ds'^2 = |(\mathbf{dr})^2\ (dx^{4'})^2| = \gamma^2 |(\mathbf{dr})^2\ (dx^4)^2|. \tag{5b}$$

Thus

$$Ds'^2 = \gamma(\alpha/2c\, x^4, r)^2 ds^2, \quad ds^2 = \gamma(-\alpha/2c\, x^4, r')^2 ds'^2. \tag{5c,d}$$

The function $\gamma$ introduces a relation between the coordinates of a source at $(r,t)$, $(r',t')$, and those of an observer at $(0,0)$, $(0,0)$ a relation that destroys the Poincare invariance of spacetime with metric $ds^2$. However, the operator $\exp(\beta_4 C_4)$ defines similarity transformations that convert relations between group generators that subsist in Minkowski spacetime into isomorphic group relations in the conformal spacetime. Thus the Poincare group reappears as an invariance group in spacetime with conformal metric $ds^2$.

Before proceeding further, we call attention to the fact that this metric is that of a spacetime which appears radially symmetric to each observer, because all may suppose themselves located at the origin of their own light cone. It is a conformal metric in the usual sense. Spatial angles between sources seen by each observer are not altered by changes of $\gamma^2$.

The spacetime generalization of angles between sources possesses a generalization of the conformal invariance exhibited by spatial angles. The metric is also distinguished by having metric coefficients $g_{ij}$, all of which rescale those of a Minkowski metric by multiplying them by the same factor, $\gamma^2$. Because of this it has many consequences that differ from those of a Robertson-Walker metric. The rescaling could allow development of consistent cosmologies without dark matter.[11a]

## 3. Experimental Determination of The Metric

For simplicity, suppose that $\alpha$ and $v/c$, $v'/c$, are small enough to use equations (4). Then

$$Dr'/dt' = dr/dt + \alpha r(1 - v^2/c^2). \tag{6a}$$

Let a spectrograph/interferometer be situated at the common origin of the r',t' and r,t coordinate systems. Suppose that EM waves have been emitted from a source with coordinates $(r_{in}, t_{in})$, $(r'_{in}, t'_{in})$ in the past, i.e., in the lower light cone. Thus $t_{in} = -|t_{in}|$, and $t'_{in} = -|t'_{in}|$. If the radial velocities v, v' of the source are directed along the outward direction of the radii, then dr/dt and dr'/dt' are positive. As a point $\rho$ at (r, t), (r', t') in the wave travels from $(r_{in}, t_{in})$, $(r'_{in}, t'_{in})$ to the origin, it has a velocity equal to c in both coordinate systems. Thus

$$r = (1 + \alpha|t|)r', \ -|t| = -|t'| - \alpha(r^2/c^2 + t^2)/2. \tag{6b}$$

From (4a-d) it follows that until this point reaches the origin, r is greater than r', and |t| is greater than |t'|. During this motion dr' is < 0, dt' > 0, and

$$dr = dr'(1 - \alpha t') - \alpha r' dt' \ , \ dt = dt'(1 - \alpha t) - \alpha r' dr'/c^2. \tag{6c,d}$$

Approximating the wavelengths $\lambda'$, $\lambda$ in the two coordinate systems by corresponding displacements |dr'|, |dr| of $\rho$, as the wave moves toward the origin one obtains from (4a-d):

$$\lambda(r,t) = \lambda'(r',t')(1 - \alpha(t' - r'/c)). \tag{6e}$$

Substituting $t' = -|t'|$ in this gives

$$\lambda(r,t) = \lambda'(r',t')(1 + \alpha(|t'| + r'/c)). \tag{6f}$$

<u>Thus, if $\alpha$ is positive, $\lambda$ is greater than $\lambda'$ at the time of emission, and the difference between the two decreases as the wave progresses.</u>

As the wave approaches the origin, $\lambda \to \lambda' \to \Lambda$, the wavelength that is measured and used to determine a Doppler shift. Let $\Lambda_{ref}$ be the relevant standard reference wavelength and set $\Delta\Lambda = \Lambda - \Lambda_{ref}$. If the velocity coordinates v', v, of the source are much less than c, then the theory of the Doppler effect implies that

$$c\,\Delta\Lambda/\Lambda_{ref} = dr'/dt' \qquad (7a)$$

is the velocity of the source in conformal spacetime. Equation (6a) then implies

$$c\,\Delta\Lambda/\Lambda_{ref} = dr/dt + \alpha\,r + O(\alpha v^2/c^2). \qquad (7b)$$

If $\alpha = 0$, one obtains the usual relation between the Doppler shift and the velocity of a source,

$$c\,\Delta\Lambda/\Lambda_{ref} = dr/dt. \qquad (7c)$$

The value of $\alpha$ can be determined if one knows the values of r and dr/dt, as well as the Doppler shift $c\,\Delta\Lambda/\Lambda_{ref}$. If $\alpha$ is not zero, the metric governing the motion of EM waves in gravity-free vacua is the conformal one, and if $\alpha$ is 0, a Minkowski metric governs their transmission.

For sources as distant as the nearest star, only Doppler shifts can be directly measured. Estimates of Hubble's constant are based on partially empirical relations between the period and brightness of Cepheid variables, and upon self-consistent estimates of values of the distances, R, to heavenly objects, and their radial velocities V.[11b] The equation

$$c\,\Delta\Lambda/\Lambda_{ref} = V + H_o\,R \qquad (8)$$

expresses Hubble's relation between these velocities and distances.[11c] In it, if the EM waves have not obviously been affected by large gravitational fields, both V and R have generally been assumed measured in a Minkowski metric.

Though Hubble's relation is primarily an empirical one presupposing a Minkowski metric, it is obviously isomorphic to the purely kinematic relation (7), which arises if the propagation of EM waves is determined by the conformal metric. The standard interpretation of Hubble's relations depends upon the standard interpretation given to Doppler shifts, which depends upon $\alpha$ being zero. If $\alpha$ is nonzero, Doppler shifts must be reinterpreted, and ($H_0\,\alpha$) must replace Ho in Hubble equations that codify motions of heavenly bodies deduced from Doppler shifts. Hill's view, and that of Tomilchik, is confirmed if $\alpha$ can be shown to have a value approximately equal to $H_0$.

The Pioneer spacecraft program produced measurements of Doppler shifts, $\Delta\Lambda/\Lambda$, of S band radar frequencies with an accuracy of ~ $10^{-12}$, and it uncovered differences between these shifts and the Doppler shifts expected from models that estimated the positions and velocities of the spaceships.[12] However, it did not directly determine these positions and velocities. The mean difference between expected and observed Doppler shifts, the "Pioneer anomaly" [12], was determined to be -2.80 x $10^{-18}$ sec$^{-1}$, to within an accuracy of better than 1. Hubble's constant,

(2.19 x 0.56) x $10^{-18}$sec$^{-1}$, has the same order of magnitude but opposite sign. Tomilchik shows that the negative sign is the expected consequence of using t instead of t' in determining the shifts in wavelength of the outgoing and incoming signals from the spacecraft - c.f. equations (6c-f) above. However, the data that produced the Pioneer anomaly is now the subject of a series of further investigations, which seem likely to significantly alter its numerical value.[13,14]

It appears that spacecraft similar to Pioneer 10 and Pioneer 11 could provide quite accurate values of all the variables required to determine α, and so determine the metric in distant regions they might visit.[14] To do so a spacecraft should, like the Pioneers, be equipped with a system to receive radar pulses from ground stations and return them as amplified pulses that have been phase locked to the received pulses. If both the Doppler shift, and the time delay between emission and receipt of successive pulses, are measured, the necessary data is provided. Repeaters installed on the outer planets and their moons might also be able to provide useful restrictions on the value of α.

## 4. Conclusion

The analysis given above establishes that the metric governing the transmission of electromagnetic waves in gravity-free vacua can be the conformal metric determined by $\exp(\beta_4 C_4) = \exp((\alpha/2c)C_4)$. The value of the group parameter fixes the metric, and restricts physical interpretations of information obtainable from electro-magnetic waves received from remote extraterrestrial sources. Minkowski's metric governs the motion of EM waves, and the Doppler effect retains its current interpretation, only if α = 0. Until the value of α is known to within a few percent, interpretations of Doppler shifts in the wavelengths of radiation received from almost all distant sources will remain seriously ambiguous. Spacecraft similar to Pioneer 10 and Pioneer 11 should be able to provide information sufficient to determine α with an accuracy greater than the accuracy to which Hubble's constant is known. If, as appears likely, it is established that α is not zero, a good deal of physics and much of cosmology, will require reexamination.

## Acknowledgements

The author is much indebted to Jerry Blakefield and Viktor Toth for their help in clarifying a number of points. He also wishes to thank Professor Lev M. Tomilchik for noticing an earlier arXiv version of the present paper and providing a preprint of his recent paper.